\documentclass[
]{ceurart}
\usepackage{float}
\usepackage{mdframed}
\usepackage{xcolor}
\usepackage{ragged2e}
\usepackage{geometry}
\usepackage{adjustbox}
\usepackage{listings}

\begin{document}

\copyrightyear{2025}
\copyrightclause{Copyright for this paper by its authors.
  Use permitted under Creative Commons License Attribution 4.0
  International (CC BY 4.0).}

\conference{
  International Workshop on AI for Access to Justice (AI4A2J 2025), Chicago, USA -- 20 June 2025.
}

\title{An AI-Based Approach to Early Reporting and Justice Initiation in Image-based Sexual Abuse. A Pilot Study.}

\author[1]{Mattia Falduti}[%
orcid=0000-0003-2078-4677,
email=mattia.falduti@thesquarecentre.org,
url=https://www.thesquarecentre.org/staff/,
]
\cormark[1]
\fnmark[1]
\address[1]{theSquare – Mediterranean Centre for Revolutionary Studies, Sant’Eufemia, 4 – Milano (MI) 20100, Italy}

\author[2]{Anca Radu}[%
orcid=0000-0003-2732-872X,
email=anca.radu@eui.eu,
url={https://www.eui.eu/people?id=anca-radu},
]
\fnmark[1]
\address[2]{European University Institute 2025, Badia Fiesolana - Via dei Roccettini 9, I-50014 San Domenico di Fiesole (FI) - Italy}

\cortext[1]{Corresponding author.}
\fntext[1]{These authors contributed equally.}

\begin{abstract}
Against the background of the widespread use of Artificial Intelligence (AI) tools in the field of justice, this paper aims to explore how an AI solution designed to draft initial reports for reporting image-based sexual abuses (IBSA) could help, support, or assist in facilitating access to justice. In our approach, access to justice is facilitated not only by easing the path to denounce IBSA (which currently has the lowest reporting rate), but also by offering an early, and thus more accurate, report draft to law enforcement authorities, providing later support also for judges. Building upon earlier approaches, we designed an improved version and tested it with three experts. In this sense, the paper advocates for AI solutions offering effective and efficient support in early reporting of IBSA.
\end{abstract}

\begin{keywords}
  Image-Based Sexual Abuse (IBSA) \sep
  AI tool \sep
  Access to Justice \sep
  Crime reporting
\end{keywords}

\maketitle

\section{Introduction}
User-generated content online can be incredibly harmful. The most diffuse example of such phenomenon is the perpetration of image-based sexual abuses, which are criminal action where perpetrators(usually ex-partners) upload real intimate or computer-generated images of their victims online. Real images can be stolen from the victim in physical settings or by hacking their devices and can become online content forever. Tremendous trauma can understandably lead survivors to withdraw and seek to forget, making the first step on the path to justice exceptionally difficult. Survivors of image-based sexual abuse (IBSA) — which includes non-consensual sharing of intimate pictures, deepfake imagery, sextortion, and upskirting — are aslo compounded by the online nature of the harm. In this context, transforming the digital environment from a source of trauma into a protective space capable of providing real, effective, and accessible help, is paramount. In other terms, such support must be easily accessible, usable, understandable, and ideally, multilingual. Crucially, it should be able to translate a survivor's initial input into a formal text suitable for both reporting abusive content to platforms and engaging with law enforcement authorities. Leveraging the significant global potential demonstrated by Generative Pre-trained Transformers (GPT) and AI solutions in recent years, we piloted and evaluated a first version of an online AI agent designed as an easy, ready-to-use tool for IBSA survivors, capable of collecting essential information and generating a formal draft report. This proof-of-concept study tests if an AI tool can serve as a practical instrument supporting survivors in their decision to report abuse and in generating the necessary initial formal text. The remainder of this paper is organized as follows: Section \ref{sec:work} introduces related work and the contribution of this study; subsequently, we describe the methodology used in Section \ref{sec:met} and the tool developed in Section \ref{sec:tool}. The outcome of a brief expert evaluation is presented in Section \ref{sec:eva}. Finally, Section \ref{sec:dis} provides a discussion of the approach and concludes the paper.

\section{Related work}\label{sec:work}
Image-based sexual abuses are a violation of a person's privacy and their human rights to dignity, sexual autonomy, and freedom of expression \cite{powell2019}, all of which occur online. To give but a few examples: i) the uploading of hacked nude photos of female celebrities shared online \cite{marwick2017}, or the automatic computer graphics generation of images or videos that portray unreal events as real through digital media manipulation (deepfake/deepnude) \cite{katarya2020}, or ii) the threat to distribute intimate materials unless a victim complies with specific demands (sextortion - a portmanteau of sex and extortion) \cite{omalley2020b}, or iii) non-consensual pornography, which is normally performed distributing intimate images previously shared under a trusted relationship.  Although such an action might be criminally relevant, most victims do not report these offences to the police, thus cybercrimes are among the least reported types of crime \cite{vandeweijer2019}. 

In particular, the practical difficulties in reporting the abusive content online have been also addressed in \cite{deangeli2021} with an expert analysis, where the authors highlighted the inhomogeneity in the practice of reporting abusive content among the most used digital platforms. More in-depth, the difficulties and the distress in reporting exactly non-consensual pornography have been analyzed and discussed in \cite{deangeli2023} , where a set of voluntaries tried to report the abuse, finding obstacles, and not performing efficiently. One of the most important issue was the fact that they were experiencing (even if simulating), a real white-page anxiety in describing the abuse. The interaction between Human Computer Interaction (HCI) and crime and the related call for joining the dots between the research community has been pointed out in \cite{bellini2020}. Recently, also the community of AI and Law highlighted the need for evaluate practical solution with users and legal experts \cite{bex2025}.

Reacting to abuse with a supportive solution has been addressed using chabots. First, a hybrid retrieval and a rule-based chatbot for survivors of IBSA has been introduced in \cite{maeng2021a} aiming at improving effectiveness against online search, emotional support, as well as the creation of a compliant letter. A second hybrid chatbot has been designed in \cite{falduti2022b} for addressing the task of completing the narrative of the abuse. 

\subsection{Contribution}
Previous chatbot approaches' for IBSA survivor support have exhibited limitations that our current work aims to address. Research such as \cite{maeng2021a} focused on a single case of IBSA and did not explore portability across languages or legal systems. The complementary expert feedback presented in \cite{falduti2022b} emphasized the critical need for reports to contain specific and unique details relevant to the justice path, stressing the importance of capturing this information early from survivors, and noting the need to contact law enforcement authorities. This study leverages these findings in order to develop an AI tool designed to handle diverse IBSA types, provide multilingual and legal system independence, and facilitate the collection of detailed, early information for report generation, thereby bridging identified gaps.

\section{Methodology}\label{sec:met}
We designed and tested our agent in a no-code online solution 
which permits giving instruction to the agent, providing a knowledge base, trying the agent, and continuously improving it.
Our goal was to provide the user with an efficient, swift and productive assistance.
The tool uses as LLM GPT 4o Mini, which permits agile performance, a multilingual adaptability, and a sufficient grade of creativity in drafting the report. In order to achieve our goal while ensuring safety for the survivors, we designed the following guidelines, instruction, and guardrails.

The initial declarative statement established the functional parameters of the digital agent, identifying it as an online support system specifically designed for individuals who have experienced image-based sexual abuse.

The scope of this tool is to provide the survivors with a formal (draft) of a report, sufficiently complete to proceed with the first steps on the justice path. This report will encompass key temporal, spatial, and contextual variables pertinent to the incident. The structure will adhere to established legal reporting standards, ensuring clarity and evidentiary integrity. While specific personal identifiers of the reporting individual will be intentionally omitted at this stage, a designated field for their inclusion will be provided, along with the date of report submission.

The data acquisition process will be sequential and focused, soliciting only the information strictly necessary for the comprehensive completion of the aforementioned report, ensuring at the same time the peculiarities of the case.

From the user experience perspective, the tool aims at maintaining a calm, respectful, and non-judgmental communication style, which is paramount throughout this interaction. Data elicitation will be conducted through discrete inquiries, individually addressing each information element  to ensure clarity and minimize potential distress.

To facilitate the content removal process, the precise Uniform Resource Locator (URL) or link containing the abusive material is required.

It is important to stress that the digital agent is programmed to refrain from requesting personally identifiable information. Instructions will be provided regarding the inclusion of such data within the designated fields of the formal report, to be subsequently furnished directly to the appropriate law enforcement agencies.

Upon completion of this interaction, the survivor will be prompted to copy and paste the generated report. They will be explicitly reminded to fill in the designated fields with their personal details and to initiate contact with law enforcement authorities, with the relevant emergency contact number (112 for Europe, 911 for the USA) reiterated for ease of access.

\section{Tool}\label{sec:tool}
Post-abuse, a timely response - including requests for content removal and the initiation of criminal investigation reports -, is recommended. Recognizing the difficulty survivors may face in undertaking these steps, our solution provides a standardized text draft. This draft enables the survivor to proactively engage with involved digital platforms and law enforcement authorities directly and simultaneously (see Fig. \ref{fig:7}).

\begin{center}
\includegraphics[width=0.8\textwidth]{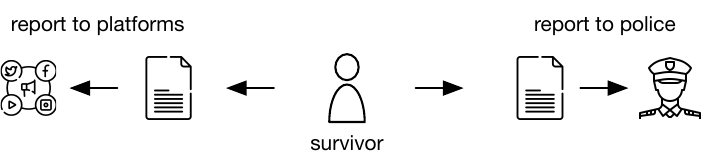}
\captionof{figure}{The usability of the same report}\label{fig:7}
\end{center}

Therefore, the tool presents an interaction with the survivor, during which they seek assistance in drafting the report. We can analyze this interaction through three key moments: agent presentation, information collection, and report generation.

\subsection{Agent Presentation}
Our priority was to clearly define the role of the agent. In line with standard chatbot conventions, the interaction begins with a greeting and the agent identifying itself as a resource for survivors of image-based sexual abuse (IBSA). To mitigate linguistic challenges and ensure accessibility, the initial message offers users the option to select their preferred language for all subsequent communication (Fig. \ref{fig:language}). Regarding language flexibility, users can modify their choice at any point until the interaction concludes, and the generated report will be provided in the selected language, with a later translation option.

\begin{center}
\includegraphics[width=0.5\textwidth]{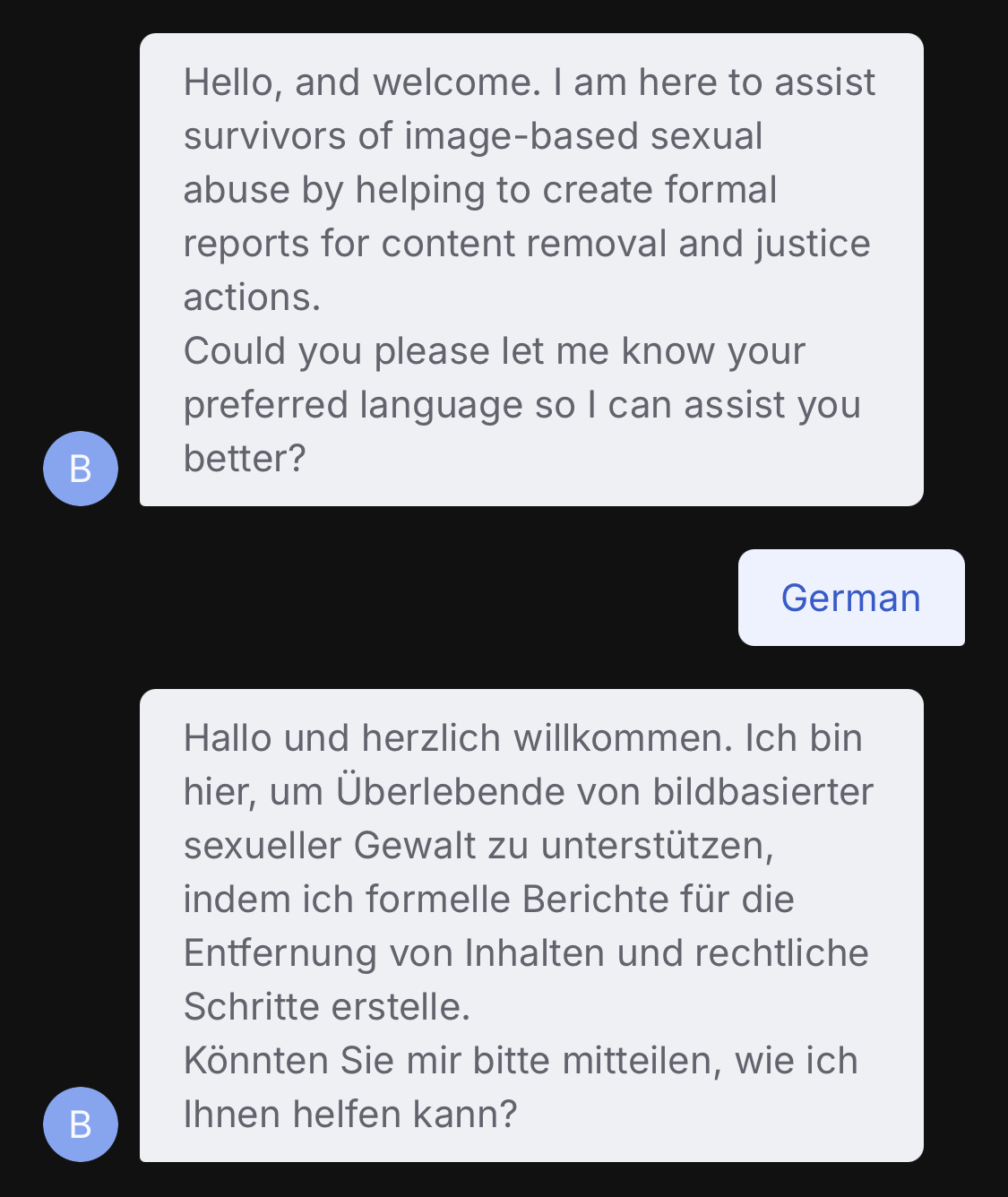}
\captionof{figure}{Presentation and language flexibility}\label{fig:language}
\end{center}

\subsection{Information collection}
Following the presentation and the selection of the language, the main activity is to collect the necessary information to create and generate the formal report draft. The interaction is designed to be concise and precise, employing simple language and short questions to avoid overwhelming users with extensive text and risking abandonment.

The agent initiates a series of questions, adapting the inquiry based on the user's responses. The primary aim at this stage is to gather information concerning at least:

\begin{itemize}
\item current location of the survivor,
\item location of the incident, 
\item time of incident occurrence,
\item time of the incident discovery,
\item brief description of the facts. 
\end{itemize}

The survivor's current location is needed to provide the correct contact information of law enforcement authorities. The incident location is necessary for the relevant justice jurisdiction, also due to the fact that the survivor's location and the incident one may differ.

\begin{center}
\begin{minipage}[b]{0.48\textwidth}
  \includegraphics[width=\linewidth]{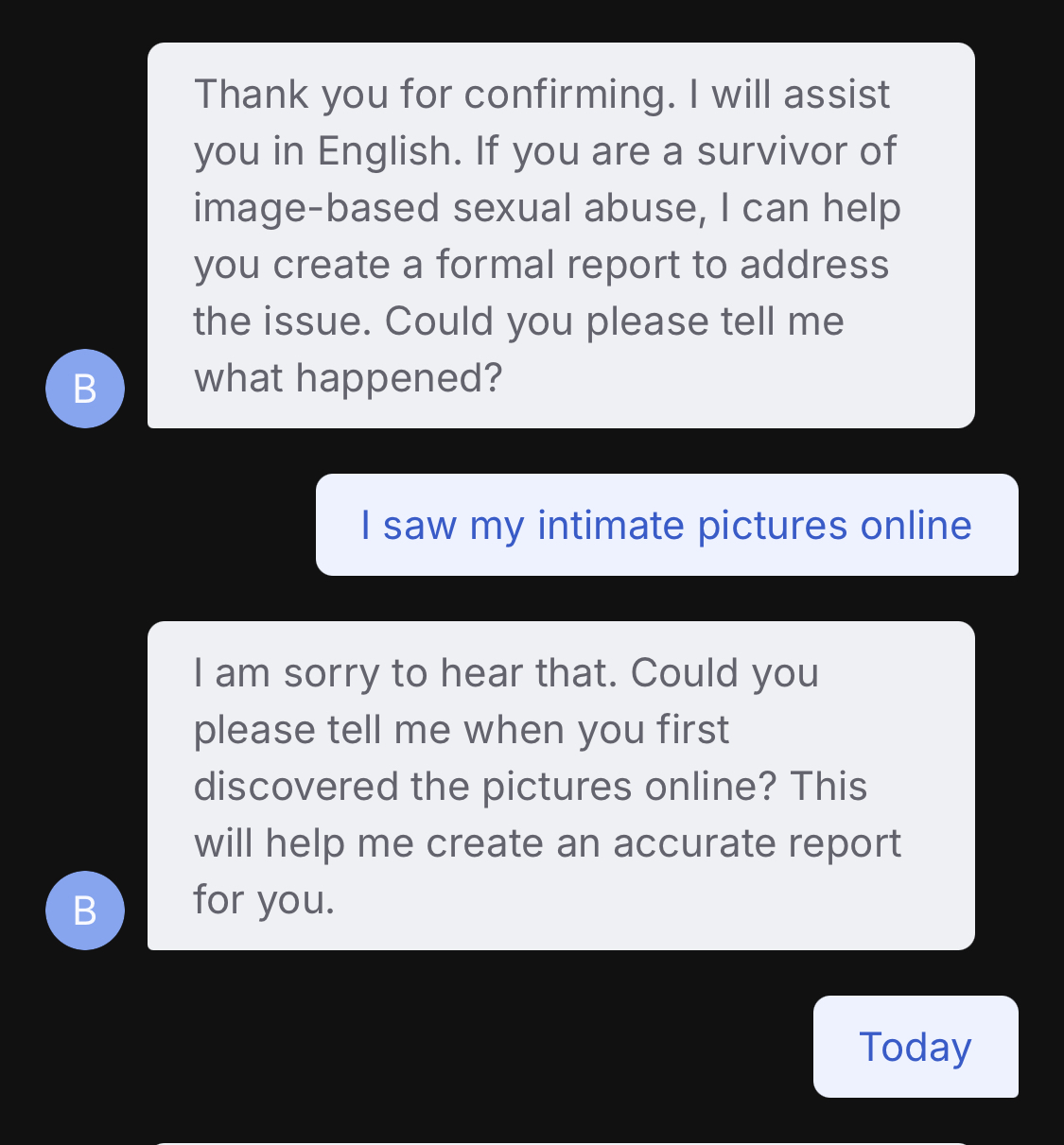}
  \captionof{figure}{Time questions}\label{fig:i1}
\end{minipage}\hfill
\begin{minipage}[b]{0.48\textwidth}
  \includegraphics[width=\linewidth]{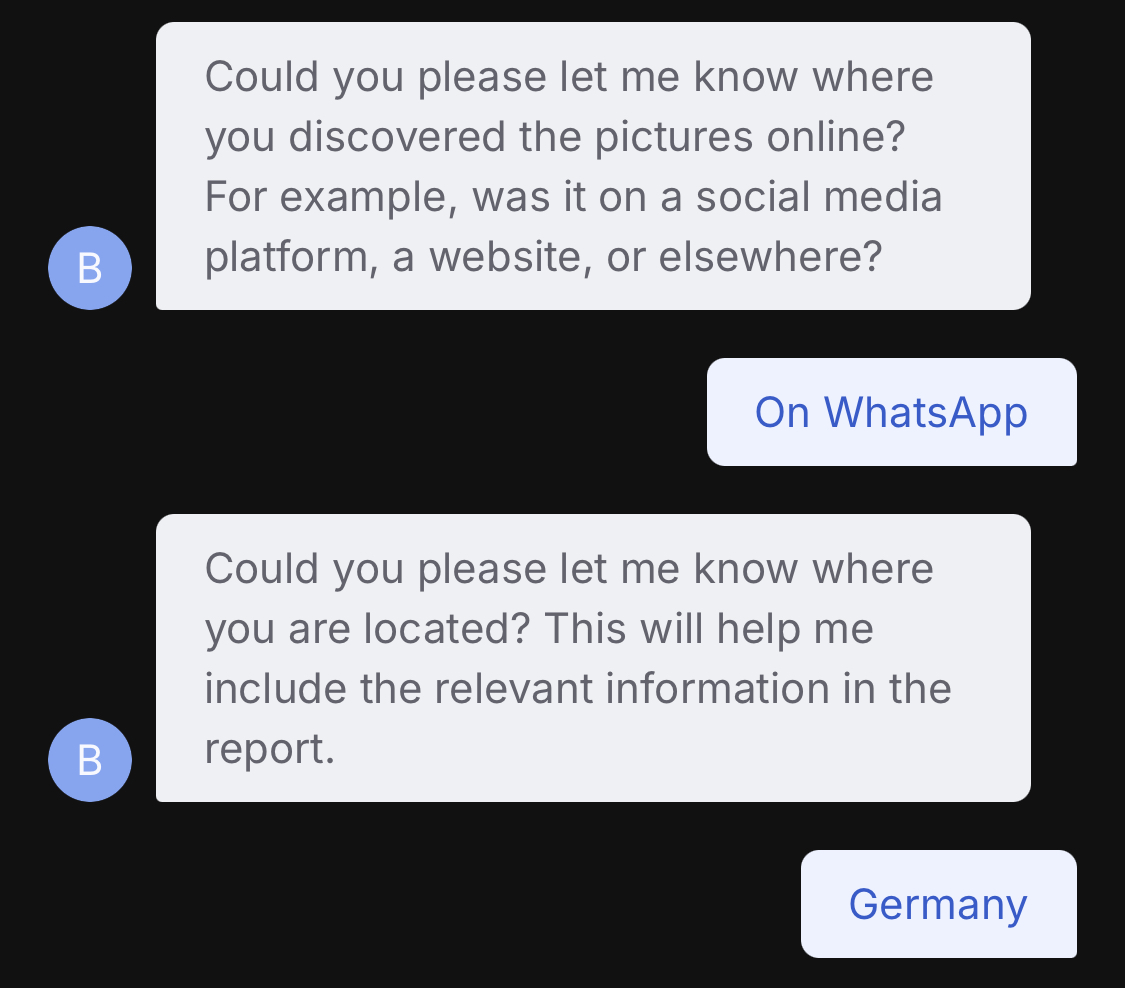}
  \captionof{figure}{Location questions}\label{fig:i2}
\end{minipage}
\end{center}

The same logic applies to the time of the incident. Unfortunately, survivors may become aware of online abuse long after it occurred.

\begin{center}
\begin{minipage}[b]{0.48\textwidth}
  \includegraphics[width=\linewidth]{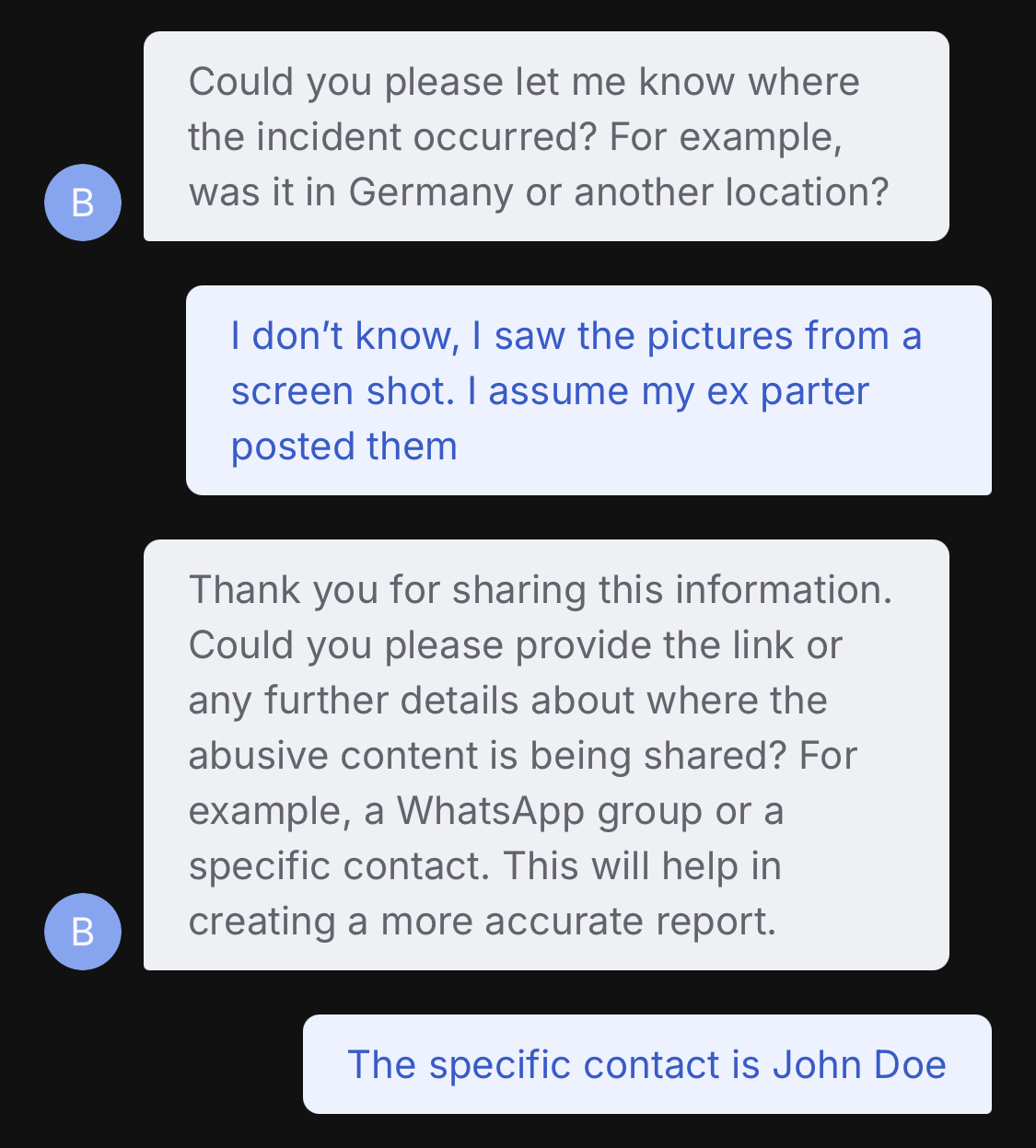}
  \captionof{figure}{Searching more details}\label{fig:i3}
\end{minipage}\hfill
\begin{minipage}[b]{0.48\textwidth}
  \includegraphics[width=\linewidth]{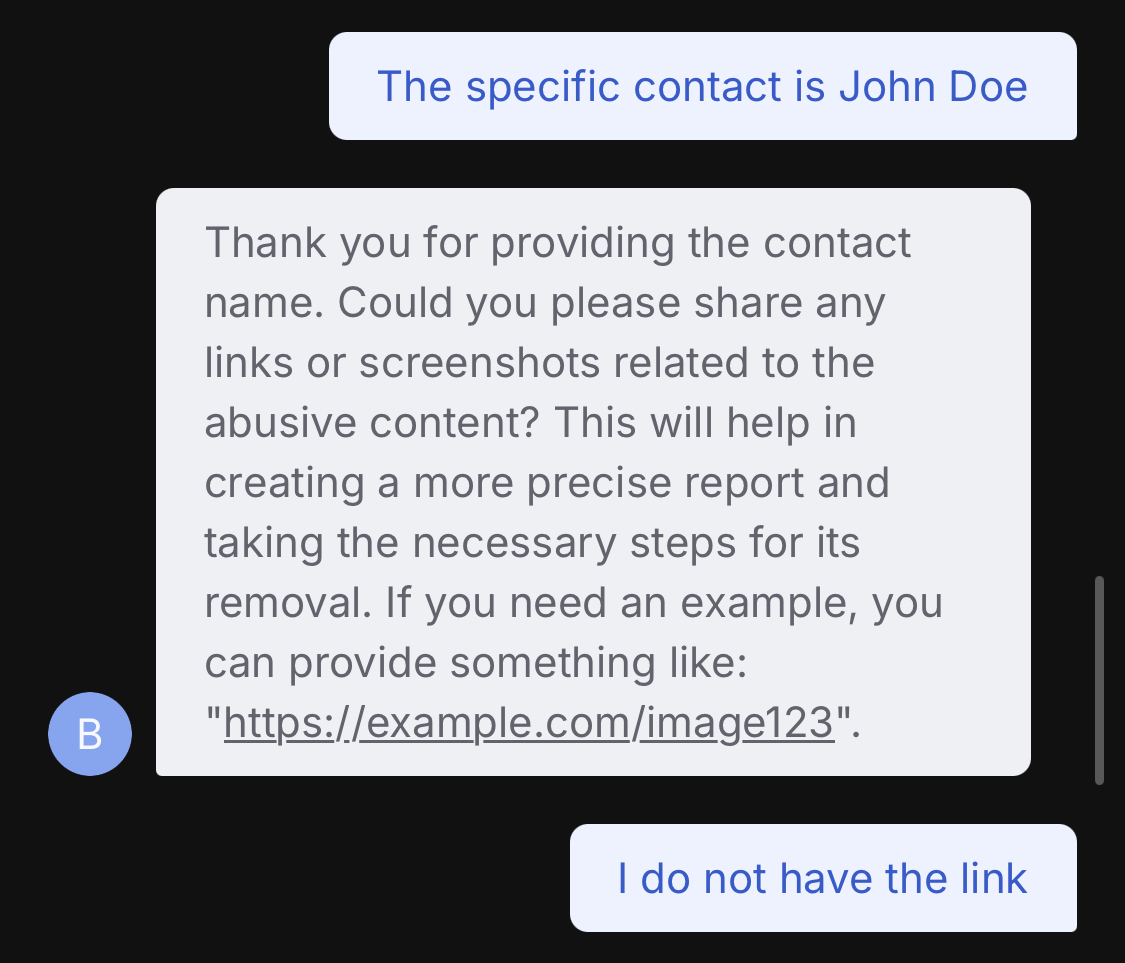}
  \captionof{figure}{Identifying the content}\label{fig:i4}
\end{minipage}
\end{center}

The agent also asks for a free description of the events and whether the survivor has access to the abusive material. This helps identify what to request for removal and what to report to the law enforcement authorities.

\subsection{Report Generation}

After collecting the necessary information for the report, the agent summarizes what the user indicated and asks for confirmation.

\begin{center}
\includegraphics[width=0.5\textwidth]{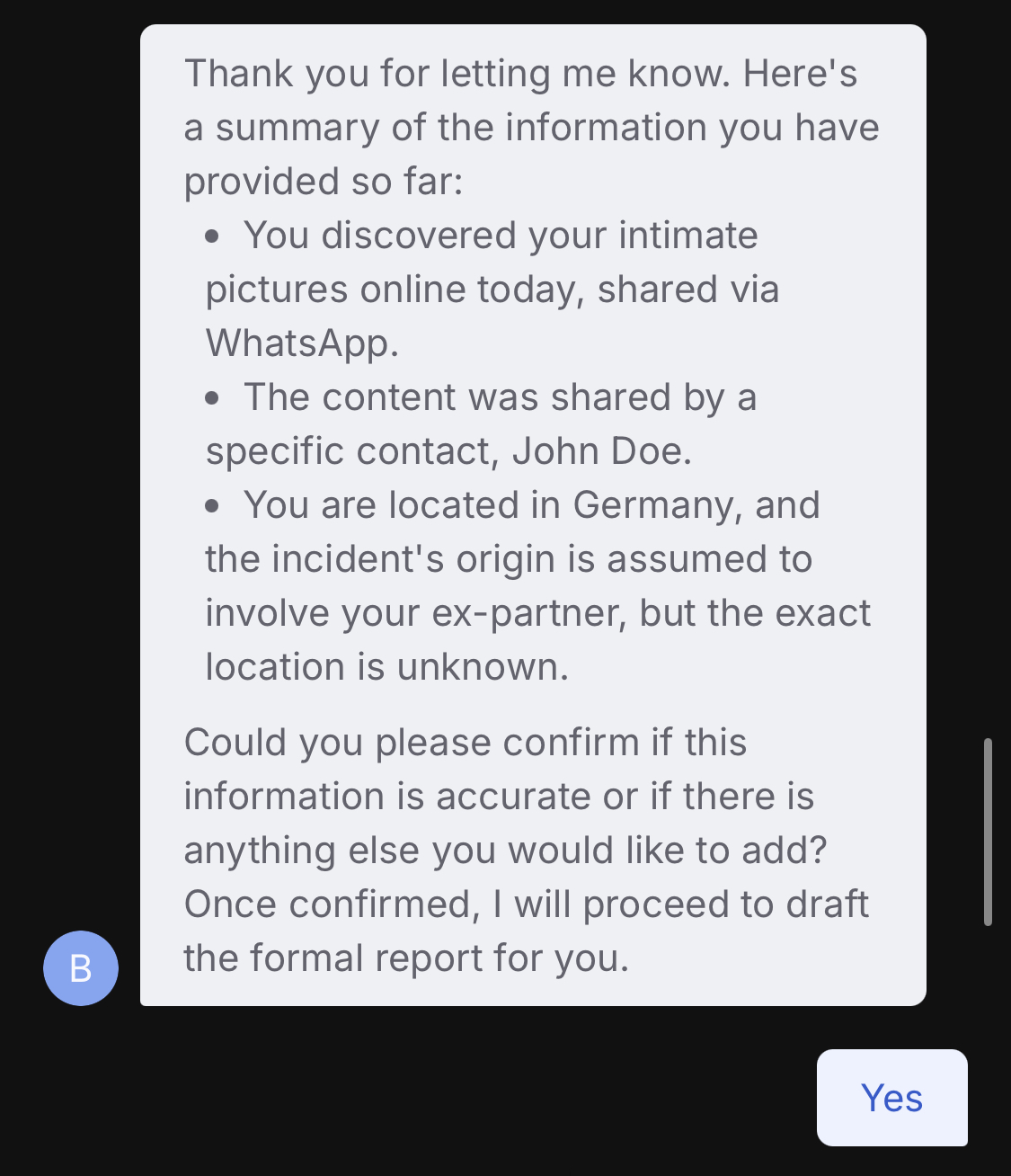}
\captionof{figure}{The brief summarization before generating the report}\label{fig:i5}
\end{center}

The last two messages are the main output. Firstly, the user is notified with the following note:\\

\textit{I will now create a formal report that you can use to report the abuse and request the removal of the content.
Please remember to complete the report with your personal data and submit it to the relevant authorities. Here is the report:}
\\

\noindent
\textbf{Formal Report on Image-Based Sexual Abuse}\\

\noindent
\textbf{Promoter of the Legal Action:} [Your Name Here]\\
\noindent
\textbf{Location of the Survivor:} Germany\\
\noindent
\textbf{Location of the Incident:} Unknown, assumed to involve the ex-partner\\
\noindent
\textbf{Date of Discovery:} 28/04/2025\\
\noindent
\textbf{Description of the Incident:} On 28/04/2025, the survivor discovered intimate pictures of themselves being shared online via WhatsApp. The content was shared by a specific contact, John Doe. The survivor suspects that their ex-partner may have posted the pictures, but the exact origin of the incident remains unknown.\\
\noindent
\textbf{Action Required:} Immediate removal of the abusive content and initiation of legal proceedings against the responsible party.\\

\noindent
The last message reports:\\
\textit{Please copy and paste this report, complete it with your personal data, and submit it to the relevant authorities. In Germany, you can contact the police by dialing \textbf{112}. If you need further assistance, feel free to ask}.

\section{Evalutation}\label{sec:eva}

For this pilot study, we engaged three legal experts and had them interact with the agent. Subsequently, we gathered their impressions and feedback. To guide their assessment, we presented them with the subsequent questions designed to evaluate three key aspects: the information collected, the effectiveness of the support provided, and the usability of the generated report.

\begin{enumerate}
    \item[(Q1)] Is the information collected by the agent sufficient to initiate the justice process?
    \item[(Q2)] Are the instructions provided to the survivor accurate and effective?
    \item[(Q3)] Can the generated text be used directly to report the content to the digital platform?
\end{enumerate}

The experts were asked to respond using a four-point Likert scale, as follows, Strongly Agree (SA), Agree (A), Disagree (D) and Strongly Disagree (SD).

\begin{center}
\begin{tabular}{c ccc}

 & Expert 1 & Expert 2 & Expert 3 \\
\midrule
\textbf{Q1} & SA & SA & SA\\
\addlinespace 
\textbf{Q2} & A & SA & SA \\
\addlinespace 
\textbf{Q3} & SA & A & SA \\
\bottomrule
\end{tabular}
\end{center}

\paragraph{Expert 1} The first expert involved has a legal background, holding a law degree (DJ) and being admitted to the Bar and worked in Court. They considered the interaction professional and supportive. The information collected was deemed sufficient to initiate the justice process, and the guidance provided to survivors was considered accurate. Furthermore, the generated text was seen as usable for reporting the content and contacting law enforcement authorities (Q1, Q3). However, the expert suggested graphically isolating the draft report and providing more precise guidance on the subsequent steps. Given the critical nature of the actions following interaction with the agent, they rated the effectiveness of the indications as poor (Q2).

\paragraph{Expert 2} The second expert involved was a member of the law enforcement authorities with 30 years of experience, 20 of them within the minor protection unit. This expert provided valuable insight from the perspective of law enforcement, particularly those who regularly listen to survivors. The feedback was again positive, with a suggestion to integrate such a solution into official websites that already offer information or allow survivors to report a crime using fillable templates requiring extensive data entry. The most appreciated aspect was the minimal time needed to complete an initial report draft, which was still considered sufficient to take the first steps in the justice process (Q1). The instructions given to survivors were also viewed positively, especially because the involvement of law enforcement authorities is unavoidable (Q2). Finally, they expressed disagreement regarding the digital platform's potential for effective engagement, even though the generated report could be useful (Q3).

\paragraph{Expert 3} Possessing a law degree (DJ) and a legal background cultivated through experience in both the Court and the public prosecutor office, this third expert indicated that the proposed solution could "set the wheels of justice in motion" (Q1). While acknowledging the accuracy of the subsequent instructions, they noted their relative simplicity, and space for improvment (Q2). The expert also appreciated the multilingual flexibility and recognized the positive aspect of the agent's ability to generate a formal report – even in its current draft form – containing sufficient information to inform the digital platform about the abuse and overcome the typical blank page anxiety when reporting using a template with open questions (Q3).

\begin{center}
\includegraphics[width=0.5\textwidth]{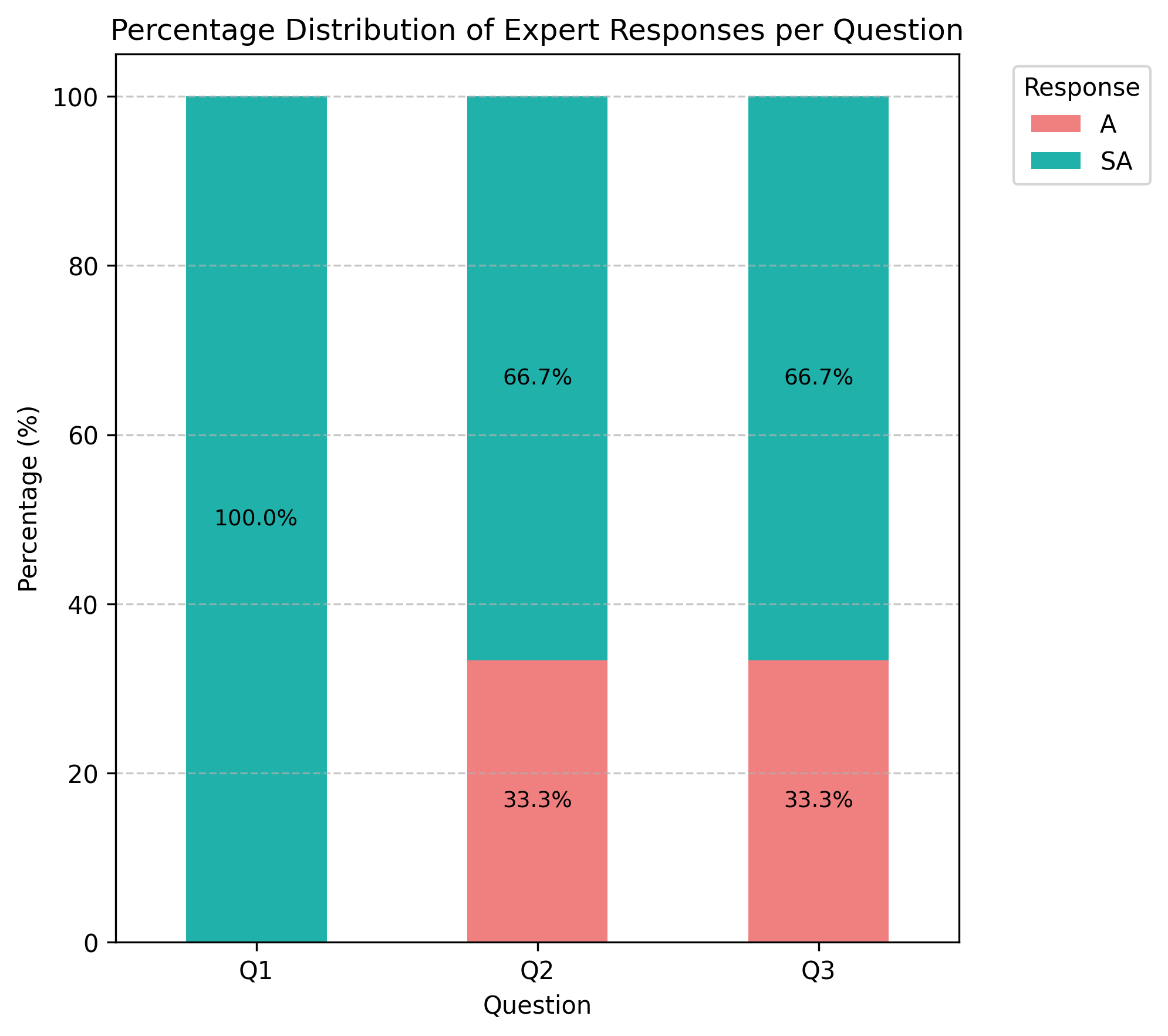}
\captionof{figure}{Responses per questions}\label{fig:i6}
\end{center}

\section{Discussion}\label{sec:dis}
We presented a pilot study of an AI-powered tool for assisting IBSA survivors in reporting abuse. The main aim was the design of a multilingual and portable tool capable of overcoming blank page anxiety often associated with the formal templates required for online content moderation or on official websites of law enforcement authorities. The involved experts confirmed that our methodological approach was promising for this purpose. Our study still has several limitations: i) the evaluation should be extended to include lay users, ii) the tool could be expanded to provide more comprehensive secondary information (such as helplines, legal aid) and iii) future work should investigate the integration of such a solution embedded within digital platforms. To conclude, we suggest that current AI solutions are able to guide survivors on the right path immediately after the abuse.


\bibliography{AI4A2J25}



\end{document}